\documentclass[numreferences]{article}

\usepackage{times}

\usepackage{amssymb}

\usepackage{epsfig}

  \renewcommand*{\Pr}{\mathop{\mathrm{Prob}}}

  \def \qedbox{\hfill\vbox{\hrule\hbox{\vrule
height1.3ex\hskip0.8ex\vrule}\hrule}}

  \newcommand{\goesto}{\rightarrow}
  \newtheorem{theorem}{Theorem}
  \newtheorem{definition}{Definition}
  \newtheorem{proposition}{Proposition}
  \newtheorem{lemma}{Lemma}
  \newtheorem{cor}{Corollary}

\def\AND{\wedge}
\def\OR{\vee}

\def\goesto{\rightarrow}
\def\implies{\Rightarrow}

\def\qed{\hfill$\Box$\newline\vspace{5mm}}

\newcommand{\eqref}[1]{Eq.~(\ref{#1})}
\newcommand{\CSP}{\mathrm{CSP}}
\newcommand{\SAT}{\mathrm{SAT}}
\newcommand{\cont}{\mathit{cont}}
\newcommand{\SATneg}{\mathrm{SAT}^\mathrm{(neg)}}
\newcommand{\Var}{\mathit{Var}}
\newcommand{\opt}{\mathit{opt}}
\newcommand{\Cl}{\mathit{Cl}}

\begin{document}

%\begin{article}

%\begin{opening} 

\title{Spines of Random Constraint Satisfaction Problems: Definition and Connection with Computational Complexity}
\author{Gabriel Istrate\footnote{istrate@lanl.gov. CCS-DSS, Los Alamos National Laboratory, Mail Stop M 997, Los Alamos, NM 87545.}, 
Stefan Boettcher\footnote{stb@physics.emory.edu. Physics Department, Emory University, Atlanta, GA 30322}, Allon G.~Percus\footnote{percus@ipam.ucla.edu. CCS-3, Los Alamos National Laboratory, Los Alamos, NM 87545
and UCLA Institute for Pure and Applied Mathematics, Los Angeles, CA
90049}} 

\maketitle
%\runningtitle{Spines of Random CSP}
%\runningauthor{Istrate, Boettcher, Percus}
%\date{}
\begin{abstract} 
We study the connection between the order of {\em phase transitions in
combinatorial problems} and the complexity of decision algorithms for
such problems.  We rigorously show that, for a class of random
constraint satisfaction problems, a limited connection between the two
phenomena indeed exists.  Specifically, we extend the definition of the
spine order parameter of Bollob\'{a}s et al. \cite{scaling:window:2sat}
to random constraint satisfaction problems, rigorously showing that for
such problems a discontinuity of the spine is associated with a
$2^{\Omega(n)}$ resolution complexity  (and thus a $2^{\Omega(n)}$
complexity of DPLL algorithms) on random instances. The two phenomena
have a common underlying cause: the emergence of ``large'' (linear size)
minimally unsatisfiable subformulas of a random formula at the
satisfiability phase transition. 

We present several further results that add weight to the intuition that
random constraint satisfaction problems with a sharp threshold and a
continuous spine are ``qualitatively similar to random 2-SAT''. Finally,
we argue that it is the spine rather than the backbone parameter whose
continuity has implications for the decision complexity of combinatorial
problems, and we provide experimental evidence that the two parameters
can behave in a different manner.  
\end{abstract} 


{\bf MR Categories:} Primary 68Q25, Secondary 82B27. 

%\end{opening} 

\section{Introduction}

The major promise of {\em phase transitions in combinatorial problems}
has been to shed light on the ``practical'' algorithmic complexity of
combinatorial problems.  A possible connection has been highlighted by
results of Monasson et al.~\cite{2+p:nature, 2+p:rsa} that are based on
experimental evidence and nonrigorous arguments from statistical
mechanics.  Studying a version of random satisfiability that
``interpolates'' between 2-SAT and 3-SAT, they suggested that the order
of the phase transition, combinatorially expressed by continuity of an
order parameter called the {\em backbone}, might have implications for
the problem's typical-case complexity.  A discontinuous or first-order
transition appeared to be symptomatic of exponential complexity, whereas
a continuous or second-order transition was correlated with polynomial
complexity.

It is understood by now that this connection is limited.  For instance,
$k$-XOR-SAT is a problem believed, based on arguments from statistical
mechanics~\cite{zecchina:kxorsat}, to have a first-order phase
transition. But it is easily solved by a polynomial algorithm, Gaussian
elimination. So, if any connection exists between first-order phase
transitions and the complexity of a given problem, it cannot involve
{\em all} polynomial time algorithms for the problem. Fortunately, this
does not end all hopes for a connection with computational complexity:
{\em descriptive complexity}~\cite{descriptive} provides a principled
way to measure the complexity of problems with respect to more limited
classes of algorithms, those expressible in a given framework.   Here we
focus on the {\em Davis-Putnam-Longman-Loveland} (DPLL) class of
algorithms~\cite{beame:dp}.

One way to identify the connection between phase transitions and
computational complexity is to formalize the underlying intuition
connecting the two notions in a purely combinatorial way, devoid of any
physics considerations.
% Indeed, there exists a nonrigorous argument of this sort:
First-order phase transitions amount to a discontinuity in the (suitably
rescaled) size of the backbone.
% But, at least
For random $k$-SAT~\cite{monasson:zecchina}, and more specifically for
the optimization problem MAX-$k$-SAT, the backbone has a combinatorial
interpretation: it is the set of literals that are ``frozen'', or assume
the same value, in all {\em optimal} assignments.  Intuitively, a large
backbone size has implications for the complexity of finding such
assignments: all literals in the backbone require specific values in
order to satisfy the formula optimally, but an algorithm assigning
variables in an iterative fashion has very few ways to know what those
``right'' values to assign are. In the case of a first-order phase
transition, the backbone of formulas just above the transition contains,
with high probability, a fraction of the literals that is bounded away
from zero. An algorithm such as DPLL that assigns values to variables
iteratively
% fashion has probably no way to consistently identify these
% ``frozen'' variables, and will
may misassign a backbone variable whose height, in a binary tree
characterizing the behavior of the algorithm, is $\Omega(n)$ where $n$
is the number of variables. This would force a backtrack on the tree.
Assuming the algorithm cannot significantly ``reduce'' the size of the
explored portion of this tree, a first-order phase transition would then
w.h.p.\ imply a $2^{\Omega(n)}$ lower bound for the running time of DPLL
on random instances located slightly above the transition.

There exists, however, a significant flaw in the heuristic argument
above: the backbone is defined with respect to {\em optimal} assignments
for the given formula,  meaning assignments that satisfy the largest
possible number of clauses (or all of them, in the case where the
formula is satisfiable).  The argument suggests that a discontinuity in
the backbone size will make it difficult for algorithms that assign
variables in an iterative manner to find {\em optimal} solutions.  The
complexity of the optimization problem is, however, often different from
that of the corresponding decision problem. For instance, that is the
case in XOR-SAT, where the decision problem is easy but the optimization
problem is hard. As mentioned above, XOR-SAT is presumed to have a
first-order phase transition, so it is not clear at all that the
continuity or discontinuity of the backbone should be the relevant
predictor for the complexity of the {\em decision} problem as well. 

Fortunately, it turns out that the intuition of the previous argument
also holds for a different order parameter, a ``weaker'' version of the
backbone called the {\em spine}, introduced
in~\cite{scaling:window:2sat} in order to prove that random 2-SAT has a
second-order phase transition. Unlike the backbone, the spine is defined
in terms of the {\em decision} problem, hence it could conceivably have
a larger impact on the complexity of these problems. Of course, the same
caveat applies as for the backbone: any connection with computational
complexity can only involve complexity classes that have weaker
expressive power than the class of polynomial time algorithms. 

We aim in this paper to provide evidence that for random constraint
satisfaction problems it is the behavior of the spine, rather than the
backbone, impacts the complexity of the underlying decision problem. To
accomplish this:

\begin{enumerate}
\item We discuss the proper definitions of the backbone and spine for
random constraint satisfaction problems (CSP).

\item We formally establish a simple connection between a discontinuity
in the relative size of the spine at the threshold and the resolution
complexity of random satisfiability problems. In a nutshell, a necessary
and sufficient condition for the existence of a discontinuity is the
existence of an $\Omega(n)$ lower bound (w.h.p.)\ on the size of
minimally unsatisfiable subformulas of a random (unsatisfiable)
subformula. But standard methods from proof
complexity~\cite{ben-sasson:resolution:width} imply that for all
problems where we can prove such an $\Omega(n)$ lower bound, there is a
$2^{\Omega(n)}$ lower bound on their resolution complexity and hence on
the complexity of DPLL algorithms as well~\cite{beame:dp}.  This
property arises from the expansion of the underlying formula's
hypergraph, and is {\em independent} of the precise definition of the
problem at hand.  Conversely we show (Theorem~\ref{second:order}) that
for {\em any generalized satisfiability problem}, a second-order phase
transition implies, in the region where most formulas are unsatisfiable,
an upper bound on resolution complexity that is smaller than any
exponential: $O(2^{\epsilon n})$ for every $\epsilon>0$.

\item We give a sufficient condition
(Theorem~\ref{sufficient:first-order}) for the existence of a
discontinuous jump in the size of the spine. We then show
(Theorem~\ref{implicates:first-order}) that this condition is fulfilled
by all problems whose constraints have no implicates of size two or
less.  Qualitatively, our results suggest that all satisfiability
problems with a continuous phase transition in the spine are
``2-SAT-like''.

\item Finally, we present experimental results that attempt to clarify
whether the backbone and the spine can behave differently at the phase
transition. The {\em graph bipartition problem} (GBP) is one case where
this seems to happen. In contrast, for random {\em 3-coloring} (3-COL),
the backbone and spine appear to have similar behavior.

\end{enumerate}

A note on the significance of our results: a first-order transition or
discontinuity in the size of the spine is weaker than a discontinuity in
the size of the backbone. In the last section of the paper we give a
numerical demonstration of an example where the backbone and spine
behave differently.  And unlike for the backbone, we do not have a
physical interpretation for the spine.  But this is not our intention.
The argument connecting the continuity of the backbone order parameter
with the complexity of decision problems is problematic, and what we
rigorously show is that --- with no physical considerations in mind ---
{\em the intuitive connection holds instead for the spine}.
% Also, in 
% the last section of the paper presents experimental work suggesting
% that {\em the backbone and the spine can behave differently}.
% For reasons of space proofs are deferred to the Appendix.

\section{Preliminaries}

Throughout this paper we assume a general familiarity with the concepts
of phase transitions in combinatorial
problems~\cite{martin:monasson:zecchina}, random
structures~\cite{bol:b:random-graphs}, and proof
complexity~\cite{beame:proof:survey}. We assume more detailed
familiarity with certain fundamental results on sharp thresholds
\cite{friedgut:k:sat,chvatal:szemeredi:resolution,ben-sasson:resolution:width},
and we make use of some of the methods associated with those results.

Two models arising in the theory of random structures are:
\begin{itemize}
\item The {\em constant probability model} $\Gamma(n,p)$. A random
string of bits $X\in\Gamma(n,p)$ is obtained by independently
setting each bit of $X$ to 1 with probability $p$, and the rest to 0.
\item The {\em counting model} $\Gamma(n,m)$. A random string
$X\in\Gamma(n,m)$ is obtained by setting $m$ bits of $X$, chosen
uniformly at random, to 1 and the rest to 0.
\end{itemize}

For the following purposes, let us work within the constant probability
model.
Consider a property $A$ that is monotonically increasing, in that
if $A$ holds for a given string of bits $X$, then changing any of these
bits from 0 to 1 preserves property $A$.  
For any $\epsilon >0$, let $p_{\epsilon}= p_{\epsilon}(n)$
be the canonical probability such that $\Pr_{X \in
\Gamma(n,p_{\epsilon}(n))}[X \mbox{ satisfies } A]= \epsilon$, where
$p_\epsilon$ increases monotonically with $\epsilon$.
{\em Sharp thresholds} are those for which the function has a
``sudden jump'' from value 0 to 1:

\begin{definition} \label{sharp}
Property $A$ has a {\em sharp threshold} iff for every $0<\epsilon
< 1/2$, we have $\lim_{n\goesto \infty} \frac{p_{1-\epsilon}(n)-
p_{\epsilon}(n)}{p_{1/2}(n)} = 0$. $A$ has {\em a coarse
threshold} if for some $\epsilon > 0$ it holds that
$\lim\inf_{n\goesto \infty} \frac{p_{1-\epsilon}(n)-
p_{\epsilon}(n)}{p_{1/2}(n)} > 0$.
\end{definition}

We will use the model of random
constraint satisfaction from Molloy~\cite{molloy-stoc2002}:

\begin{definition}\label{model} Let ${\cal D} = \{0,1,\ldots, t-1\}$,
$t\geq 2$ be a fixed set. Consider all $2^{t^{k}}-1$ possible 
nonempty sets of $k$-ary constraint templates (relations) 
 with values taken from ${\cal D}$. Let
${\cal C}$ be such a nonempty set of constraint templates.

A random formula $\phi\in\CSP({\cal C})$ is a set of constraints 
specified under the counting model by the following procedure:

\begin{enumerate}
% \item $n$ is the number of variables.
% 
% \item $m$ is the number of clauses, chosen by the following
% procedure: first
\item Select, uniformly at random and with replacement,
$m$ hyperedges of the complete $k$-uniform hypergraph on $n$ variables.

\item For each hyperedge, choose a random ordering of the variables
involved in it. Choose a random constraint template from  ${\cal C}$ and 
apply it to the list of (ordered) variables.
\end{enumerate}

We use the notation $\SAT({\cal C})$ (instead of $\CSP({\cal C})$)
when t=2.
% Also, for $\Phi$ an instance of $\CSP({\cal C})$ we
% denote by $Opt(\Phi)$ the set of {\em optimal assignments} for $\Phi$, and by $opt(\Phi)$ the number of constraints left
% unsatisfied by an optimal assignment. 
\end{definition}

For an instance $\Phi\in\CSP({\cal C})$, we denote by $\Var(\Phi)$ the
set of variables that actually appear in $\Phi$, and by $\opt(\Phi)$ the
number of constraints left unsatisfied by an {\em optimal assignment}
for $\Phi$.
% Finally, we denote by $\Opt(\Phi)$ the set of
% {\em optimal assignments} for $\Phi$, and by $\opt(\Phi)$ the number of
% constraints left unsatisfied by an optimal assignment. 

% We will denote by $Var$ the set of variables a particular instance 
% of $\CSP({\cal C})$ is {\em defined on}, and by $Var(\Phi)$ the variables 
% that {\em effectively appear in $\Phi$}. 
% When dealing with {\em boolean} constraint satisfaction problems, 
% we will also use $Lit(\Phi)$ to denote the set of {\em literals} 
% (variables and negated variables) the formula is defined on.

Just as in random graphs~\cite{bol:b:random-graphs}, under fairly
liberal conditions one can use the constant probability model
instead of the counting model from the previous definition. The
interesting range of the parameter $m$ is when the ratio $m/n$ is a
constant, $c$, called the {\em constraint density}. The original
investigation of the order of the phase transition in $k$-SAT used an
order parameter called {\em the backbone},
\begin{equation}\label{bb:initial}
B(\Phi) = \{ x\in \Var(\Phi)\ |\ \exists\ W\in \{x,\overline{x}\} : 
\opt(\Phi \cup W) >\opt(\Phi) \},
% B(\Phi) = \{ x\in \Var(\Phi)\ |\ \exists\ \lambda \in \{0,1\} : 
% \forall\ X \in \Opt(\Phi), X(x)=\lambda \},
\end{equation}
or more precisely {\em the backbone fraction} 
\begin{equation}\label{fb}
f_{B}(\Phi)=\frac{|B(\Phi)|}{n}.
\end{equation} 

% One could define the backbone in term of variables, rather than literals. 
% However, to compute the backbone fraction we would have to normalize by $|Var(\Phi)|$ instead of $|Var(\Phi)|$, and the definition would not change the behavior of the parameter.  

Bollob\'{a}s et al.~\cite{scaling:window:2sat} have investigated
the order of the
phase transition in $k$-SAT (for $k=2$) under a different order
parameter, a ``monotonic version'' of the backbone called {\em the
spine}
\begin{equation}\label{spine:initial}
S(\Phi) = \{ x\in \Var(\Phi) \ |\ \exists\  W\in\{x,\overline{x}\},
\Xi \subseteq \Phi : \Xi \in
\SAT, \Xi \AND W \in \overline{\SAT}\}.
\end{equation}
Here, ``$\in\SAT$'' means ``is satisfiable'' and
``$\in\overline{\SAT}$'' means ``is unsatisfiable''.

The corresponding version of \eqref{fb} is

\begin{equation}\label{fspine}
f_{S}(\Phi)=\frac{|S(\Phi)|}{n}.
\end{equation} 

They showed that random 2-SAT has a continuous (second-order)
phase transition: the size of $f_{S}$ approaches zero w.h.p.\
(as $n\goesto
\infty$) for constraint density $c<c_\mathrm{2-SAT}=1$, and is continuous at
$c=c_\mathrm{2-SAT}$.
By contrast, nonrigorous arguments from statistical
mechanics~\cite{monasson:zecchina} imply that for 3-SAT the parameter 
$f_{B}$ jumps discontinuously from zero to a positive value at the
transition point $c=c_\mathrm{3-SAT}$ (a first-order phase transition).

\section{How to define the backbone/spine for random CSP (and beyond)}

We would like to extend the concepts of backbone and spine to
general constraint satisfaction problems.
The extended definitions
must preserve as many of the properties of the
backbone/spine as possible. 

 Certain
differences between the case of random $k$-SAT and the general
case force us to employ an alternative definition of the
backbone/spine. The most obvious is that
\eqref{spine:initial} involves negations of variables, unlike Molloy's
model. Also, these definitions are inadequate
for problems whose solution space presents a relabeling symmetry,
such as the case of {\em graph coloring} where
the set of (optimal) colorings is closed under permutations of the
colors. Due to this symmetry, no variable can be ``frozen'' to a fixed
value $\lambda$ as in \eqref{bb:initial}.

We therefore define the backbone/spine of a
random instance of $\CSP({\cal C})$ in a slightly different manner. 
Let $\hat{C}$ be the set of constraints obtained by applying the
constraint templates in ${\cal C}$ to all ordered lists of $k$ variables
chosen from the set of all $n$ variables.

\begin{definition}\label{spine-first} 
\[
B(\Phi) = \{x\in \Var(\Phi)\ |\ \exists\ C \in \hat{C} : x\in C,
\opt(\Phi \cup C) >\opt(\Phi) \},
\]
\[
S(\Phi) = \{ x\in \Var(\Phi)\ |\ \exists\ C\in \hat{C}, \Xi \subseteq \Phi : x\in C, \Xi \in \CSP,
\Xi \cup C \in \overline{\CSP}\}.
\]

\end{definition}

For $k$-CNF formulas whose (original)
backbone/spine contains at least three literals, a variable $x$ is
in the (new version of the) backbone/spine if and only if either
$x$ or $\overline{x}$ were present in the old version. In
particular the new definition does not change the order of the
phase transition of random $k$-SAT.

Alternatively, in studying 3-colorability (3-COL) of random graphs
$G=(V,E)$, Culberson and Gent~\cite{frozen:development} defined the
spine of a colorable graph $G$ to be the set of vertex pairs $(x,y)\in
V^{2}$ that get assigned the same color in {\em all} colorings of $G$. 

Following up on the idea of defining the backbone and spine in terms of {\em constraints} rather than {\em variables}, and by analogy with the definition in~\cite{scaling:window:2sat},  
one can extend the definition of $S(G)$ to general graphs\footnote{Culberson and Gent employ an ``effective'' version of the spine they call {\em frozen development} that is more amenable to 
experimental analysis. Frozen development is a subset of the spine, as
defined in \eqref{spine-3col}.} by 

\begin{equation}\label{spine-3col}  
S(G) = \{ (x,y)\in V^{2}\ |\ \exists\ H \subseteq G : H \in
\mbox{3-COL}, H \cup (x,y) \in \overline{\mbox{3-COL}}\}.
\end{equation}

We can further extend these definitions to all random constraint satisfaction 
problems $\CSP({\cal C})$: 

\begin{definition}\label{spine-two} 
\[
B_{C}(\Phi) = \{C \in \hat{C}\ |\ \opt(\Phi \cup C) > \opt(\Phi) \},
\]
\[
S_{C}(\Phi) = \{ C\in \hat{C}\ |\ \exists\ \Xi \subseteq \Phi : \Xi \in
\CSP, \Xi \cup C \in \overline{\CSP}\}.
\]
\end{definition} 
Similarly,
% if we denote by ${\cal C}(\Phi)$ the set of constraints obtained 
% by applying the constraint templates in ${\cal C}$ to all ordered lists 
% of $k$ variables chosen from $Var$, 
one can define the {\em backbone/spine fraction} by 
\[
f_{B_C}(\Phi)= \frac{|B_{C}(\Phi)|}{|\hat{C}|},  
\]
and 
\[
f_{S_C}(\Phi)= \frac{|S_{C}(\Phi)|}{|\hat{C}|}.
\]

We will refer to these concepts as the {\em constraint-based}
backbone/spine (fractions), as opposed to the previously defined {\em
variable-based} quantities. The two are clearly related.  For instance 
one can easily show that 
\[
B(\Phi)=\cup_{C\in B_{C}(\Phi)} \Var(C),
\]
where $\Var(C)$ represents all variables appearing in constraint $C$
alone.  It is also clear that $|B_{C}(\Phi)|=O(|B(\Phi)|^k)$ and
similarly for the spine.  Since $|\hat{C}|=\Theta(n^k)$, it follows that
the continuity of $f_{B}$ or $f_{S}$ implies
the continuity of $f_{B_C}$ or $f_{S_C}$.  However, the converse is
not in general true, and so
% 
% (a similar identity holds for the two definitions of the spine). 
% 
% If the variable-based backbone/spine of an 
% instance of $\CSP({\cal C})$ has size $u$, then the constraint-based 
% backbone/spine has size $O(u^{k})$. It follows readily 
% that continuity of $f_{B}$ ($f_{S}$) will imply 
% the continuity of $f_{B_C}$ ($f_{S_C}$). 
% 
% On the other hand, while the size of the variable-based 
% backbone is at most $k$ times the size the relation of the constraint-based 
% backbone (and similarly for the spine), this relation is {\em not} powerful 
% enough to show that
the two backbone/spine fractions do not necessarily behave in the same 
way.
% The reason for that is that we are using different normalization factors 
% ($|Var|$ in one case, $\theta(|Var|^{k})$ in the other) 
% when defining the two versions of the order parameters. 

Given the two types of definitions, which
should we choose? The answer depends on the problem, as well as 
on the issue we wish to address. For instance,
in the statistical mechanics analysis of
combinatorial problems, the presumably ``correct'' definition of the backbone
emerges from the analysis
undertaken in \cite{monasson:zecchina} for random $k$-SAT. 
But since we are interested
in a combinatorial definition, with no physics considerations in
mind, the only principled way to choose between the two types of
order parameters (one based on variables, the other based on
constraints) is based on the class of algorithms we are concerned
with. In the case of random constraint satisfaction problems and DPLL
algorithms, it is variables that get assigned values, so
Definition~\ref{spine-first} is preferred. On the other hand, 
constraint-based definitions can make sense for problems that
share some characteristics with random 3-COL (i.e., binary
constraint satisfaction problems, and problems with built-in
symmetries of the solution space). In a later section we will see an
example, the case of {\em graph bipartition}, where the
constraint-based backbone and spine seem to behave differently.
(Whether one can come with a natural example of this phenomenon
for the variable-based backbone is an interesting open problem.)

\section{Spine discontinuity and resolution complexity of random CSP}\label{main:section}

In this section we will study the continuity of the spine-based order
parameter  $f_{S}$ for {\em boolean} random constraint satisfaction,
or satisfiability, problems. The kind of continuous/discontinuous
behavior we are looking for is formalized by the following definition (a
similar one can be given for the constraint-based versions of the order
parameter):   

\begin{definition}\label{cont} 
 Let ${\cal C}$ be such that $\SAT({\cal C})$ has a
sharp threshold. Problem $\SAT({\cal C})$ has a {\em discontinuous spine}
if there exists $\eta > 0$ such that for every sequence $m = m(n)$ we have
\begin{equation}\label{jump:general}
\lim_{n\goesto \infty} \Pr_{m=m(n)}[\Phi \in \SAT] = 0 \implies
\lim_{n\goesto \infty}\Pr_{m=m(n)}[ f_{S}(\Phi)\geq
\eta]= 1.
\end{equation}
If, on the other hand, for every $\epsilon >0$ there exists a constant 
$c=c({\epsilon})$ such that the map $\epsilon \goesto c({\epsilon})$ is monotonically increasing and 
\begin{equation}\label{jump:continuous}
\lim_{n\goesto \infty} \Pr_{m=c({\epsilon})n}[\Phi \in \SAT] = 0\
\mathrm{and}\ \lim_{n\goesto \infty}\Pr_{m=c({\epsilon})n}[
f_{S}(\Phi)\geq \epsilon]= 0
\end{equation}
we say that $\SAT({\cal C})$ has a {\em continuous spine}.
\end{definition}

We now give a simple observation that will be the basis for identifying 
discontinuities of the spine: 

\begin{proposition}\label{spine:unsat}
Let $\Phi$ be a minimally unsatisfiable formula, and let $x$ be a
literal that appears in $\Phi$. Then, by Definition~\ref{spine-first}, $x\in S(\Phi)$.
\end{proposition}

\begin{proof}
There exists $C\in \Phi$ such that $x\in C$. But $\Phi \setminus C$ is satisfiable and 
$(\Phi \setminus C)\cup C$ is not, 
thus $x\in S(\Phi)$.   
\end{proof} 
\qed

\begin{cor}\label{3sat:first-order}
$k$-SAT, $k\geq 3$ has a discontinuous spine.
\end{cor}

\begin{proof}
To show a discontinuous spine it is sufficient to show
that a random unsatisfiable formula contains w.h.p.\ a minimally
unsatisfiable subformula involving a linear number of literals.
% A way to accomplish this is by directly employing the
In the Chv\'{a}tal-Szemer\'{e}di proof
\cite{chvatal:szemeredi:resolution} that w.h.p.\ random $k$-SAT has
exponential resolution size for $k\geq 3$, % where the above claim
the claim is implicitly proved. 
\end{proof}
\qed

\begin{definition} The {\em width} of a resolution proof $P$ of the unsatisfiability of a CNF-formula $F$ is defined to be the maximum number of 
literals in any clause that appears in the proof $P$.   

If $\Phi$ is an instance of $\SAT({\cal C})$, denote by $\Cl(\Phi)$ the
CNF formula obtained by expressing each constraint of $\Phi$ as a
conjunction of {\em clauses} (i.e., expressing $\Phi$ in conjunctive
form).  

The {\em resolution complexity} of an instance $\Phi$ of
$\SAT({\cal C})$ is defined as the length of the smallest resolution 
proof  of $\Cl(\Phi)$. 
\end{definition} 

 A simple observation is that a continuous spine
has implications for resolution
complexity:

\begin{theorem}\label{second:order}
Let ${\cal C}$ be a set of constraint templates such that $\SAT({\cal C})$ has a
continuous spine.
Then for every constraint density $c>lim_{\epsilon \goesto
0} c({\epsilon})$, and {\em every} $\epsilon>0$, random
formulas of constraint density $c$ have w.h.p.\ resolution
complexity $O(2^{\epsilon n})$.
\end{theorem}

\begin{proof}

Because of Proposition~\ref{spine:unsat} and the fact that  
$\SAT({\cal C})$ has a continuous spine, for every $\epsilon>0$, minimally
unsatisfiable subformulas of a random formula $\Phi$ with
constraint density $c({\epsilon})$ contain w.h.p.\ at most $\epsilon n$ 
variables. Consider the backtrack tree of the natural DPLL algorithm that
tries to satisfy constraints one at a time on such a minimally
unsatisfiable subformula $F$. By the usual correspondence between
DPLL refutations and resolution complexity (e.g., \cite{beame:dp})
this yields a resolution proof of the unsatisfiability of $\Phi$
having size at most $2^{\epsilon n}$.

Taking $\epsilon$ to be small enough that $c({\epsilon})<c$, and using the 
fact that resolution complexity of a random formula is a monotonically
decreasing function of the constraint density, we get the desired result.  
\end{proof}
\qed

Let us observe that we have stated the preceding theorem using condition
$c>lim_{\epsilon \goesto 0} c({\epsilon})$ since we cannot be sure, even
for $k$-SAT, that the phase transition takes place at a constant value
of the constraint density $c$. In practice one would of course expect
that, for a problem with a continuous spine, there
exists a sequence $c({\epsilon})$ as in Definition~\ref{cont} having the
constraint density at the phase transition as its limit.

\begin{definition}
Denote by $|F|$ the
number of constraints that appear in formula $F$. Define  

\[ c^{*}(F)= \max\left\{
\frac{|H|}{|\Var(H)|}: \emptyset \neq H \subseteq
F\right\}.
\]
\end{definition}

The next result gives a sufficient condition for a generalized
satisfiability problem to have a discontinuous spine. Interestingly, it is one condition
studied in \cite{molloy-stoc2002}.

\begin{theorem} \label{sufficient:first-order}
Let ${\cal C}$ be such that $\SAT({\cal C})$ has a
sharp threshold. If there exists $\epsilon > 0$ such that for
every minimally unsatisfiable formula $F$ it holds that
$c^{*}(F) > \frac{1+\epsilon}{k-1}$, 
then $\SAT({\cal C})$ has a discontinuous spine.

\end{theorem}

\begin{proof}
The proof is similary to that of Corollary~\ref{3sat:first-order}: we will show that 
w.h.p.\ a random formula contains a minimally unsatisfiable subformula containing a linear number of variables, and apply Proposition~\ref{spine:unsat}. 

To accomplish that, we first recall the following concept from
\cite{chvatal:szemeredi:resolution}:

\begin{definition}
Let $x,y>0$. A $k$-uniform hypergraph with $n$ vertices is {\em
($x$,$y$)-sparse} if every set of $s\leq xn$ vertices contains at
most $ys$ edges.
\end{definition}

We also recall Lemma 1 from the same paper.
\begin{lemma}\label{sparsity:hypergraph}
Let $k,c>0$ and $y>1/(k-1)$. Then w.h.p.\ the
random $k$-uniform hypergraph with $n$ vertices and $cn$ edges is
$(x,y)$-sparse, where
\begin{equation}\label{x-sparse}
x = \left( \frac{1}{2e}\left(\frac{y}{ce}\right)^{y}\right)^{\frac{1}{y(k-1)-1}}.
\end{equation}
\end{lemma}

Let $y=\frac{1+\epsilon}{k-1}$.  Directly applying
Lemma~\ref{sparsity:hypergraph}, w.h.p.\ a random
$k$-uniform hypergraph with $cn$ edges is $(x_{0},y)$ sparse, for
$x_{0}=(\frac{1}{2e}(\frac{y}{ce})^{y})^{\frac{1}{\epsilon}}$.
The critical observation is then that the existence of a minimally
unsatisfiable formula with $xn$ variables and with $c^{*}(F) >
\frac{1+\epsilon}{k-1}$ implies that the $k$-uniform hypergraph
associated with the given formula is {\em  not} $(x,y)$-sparse.
It follows that any formula
with fewer than $x_{0}n/k$ constraints (and thus fewer than 
 $x_{0}n$ variables) is satisfiable. Therefore, any 
minimally unsatisfiable subformula of random formula $\Phi$ has more
than $x_{0}n/k$ constraints. 

To show that such formulas have many variables, we again employ the expansion 
of the formula hypergraph given by Lemma~\ref{sparsity:hypergraph}, and 
infer that {\em all} subformulas of size less than $xn$ of $\Phi$ (in particular those that are also subformulas of a minimally unsatisfiable subformula of 
$\Phi$) have a linear number of variables. 
\end{proof}
\qed

One can give an explicitly defined class of satisfiability
problems for which the previous result applies:

\begin{theorem}\label{implicates:first-order}
Let $k\geq 2$ and let ${\cal C}$ be such that $\SAT({\cal C})$ has a
sharp threshold.  If  {\em no} clause template $C\in {\cal C}$ has (when
expressed as a CNF-formula) an implicate of length 2 or 1 then
\begin{enumerate}
\item For every minimally unsatisfiable formula $F$, 
$
c^{*}(F)\geq \frac{2}{2k-3}$. Therefore $\SAT({\cal C})$ satisfies the conditions of the previous
theorem, i.e., it has a discontinuous spine.
\item Moreover, there exists a constant $\eta >0$ such that w.h.p.\ 
random instances of $\SAT({\cal C})$ have $\Omega(2^{\eta n})$ resolution
complexity\footnote{This result subsumes some of the results in
\cite{mitchell:cp02}. While a preliminary version of this paper was under consideration (and publicly available \cite{istrate:allthat}) related and technically more sophisticated results have been given independently in  \cite{molloy:focs2003}.}.
\end{enumerate}
\end{theorem}

The condition in the theorem is violated, as expected, by random 2-SAT.
It is also violated by the random version of the NP-complete problem
1-in-$k$-SAT. This can be seen as follows.  The problem can be
represented as $\CSP({\cal C})$, for ${\cal C}$ a set of $2^{k}$
constraints corresponding to all ways to negate some of the variables,
and has a rigorously determined ``2-SAT-like'' location of the
transition point~\cite{istrate:1ink:sat}. However, the formula 
\[
C(x_{1}, x_{2}, \ldots, x_{k-1},
x_{k})\AND C(\overline{x_{k}}, x_{k+1},  \ldots , x_{2k-2},
x_{1})
\]

\[
\AND C(\overline{x_{1}}, x_{2k-1}, \ldots, x_{3k-3},
\overline{x_{k}}) \AND C(x_{k}, x_{3k-2}, \ldots, x_{4k-4},
\overline{x_{1}}),
\]
where $C$ is the constraint ``1-in-$k$'', is minimally
unsatisfiable but has clause/variable ratio $1/(k-1)$ and
implicates $\overline{x_{1}} \OR \overline{x_{k}}$ and $x_{1}\OR
x_{k}$.

\vspace{5mm}

\begin{proof}

\begin{enumerate}
\item For any real $r \geq 1$, formula $F$ and set of clauses
$G\subseteq F$,
 define the {\em $r$-deficiency of $G$}, $\delta_{r}(G)=
r|G|-2|\Var(G)|$. Also define
\begin{equation} \label{max}
\delta^{*}_{r}(F)= \max\{\delta_{r}(G): \emptyset \neq G \subseteq
F\}
\end{equation}

\begin{definition}
Let $F$ be a formula, and let $C_{1}, C_{2}, \ldots, C_{i}, \ldots, C_{m}$ be 
a listing of the constraints in $F$. 
\item A variable $v$ is {\em private} for constraint $C_{i}$ if $v$
appears in $C_{i}$ but in no other constraint.
\item Variable $v$ is {\em free in $C_{i}$} if $v$ appears in $C_{i}$ but in no $C_{j}$, $j<i$. Otherwise we say that $v$ is {\em bound in $C_{i}$}.  
\end{definition}

We claim that for any minimally unsatisfiable $F$,
$\delta^{*}_{2k-3}(F)\geq 0$. Indeed, assume this was not true.
Then there exists such $F$ such that:
\begin{equation}\label{deff}
\delta_{2k-3}(G)\leq -1\mbox{ for all }\emptyset \neq G\subseteq
F.
\end{equation}
\begin{lemma} \label{1-transversal}
Let $F$ be a formula for which condition~\ref{deff} holds. Then
there exists an ordering $C_{1}, \ldots, C_{|F|}$ of constraints
in $F$ such that each constraint $C_{i}$ contains at least $k-2$
variables that are free in $C_{i}$. 
\end{lemma}

\begin{proof}
Denote by $v_{i}$ the number of variables that appear in {\em exactly}
$i$ constraints of $F$. We have $ \sum_{i\geq 1} i v_{i} = k |F|$,
therefore $2|\Var(F)|-v_{1}\leq k |F|$. This can be rewritten as
$v_{1}\geq 2|\Var(F)|-k|F|> |F| (2k-3 - k)= (k-3) |F|$, where we use
\eqref{deff}. Therefore there exists at least one constraint $C$ in $F$
with at least $k-2$ variables that are private in $F$, hence necessarily
free in $F$.  We set $C_{|F|}=C$ and apply this argument recursively to
$F\setminus C$.
\end{proof}
\qed

Let us show now that $F$ cannot be minimally unsatisfiable. Construct a
satisfying assignment for $F$ incrementally, so that the partial assignment 
constructed up to stage $j$ will satisfy constraints $C_{1}, \ldots, C_{j}$. 

Indeed, suppose we have constructed a partial assignment that satisfies 
$C_{1}, \ldots, C_{j-1}$, and consider now constraint
$C_{j}$. At most two of the variables in $C_{j}$ are bound in
$C_{j}$. Since $C_{j}$ has no implicates of length two or less, no matter 
what the assignment to these two variables might have been in the previous 
stages, one can set the variables that are free in $C_{j}$ in a way that 
satisfies this clause. Iteratively performing this construction 
yields a satisfying assignment for $F$, in contradiction with our
assumption that $F$ was minimally unsatisfiable.

Therefore $\delta^{*}_{2k-3}(F)\geq 0$, a statement equivalent to
our conclusion.

\item To prove the resolution complexity lower bound we use the
size-width connection for resolution complexity obtained in
\cite{ben-sasson:resolution:width}: it is sufficient to prove that 
there exists $\eta >0$ such that w.h.p.\ random instances of $\SAT({\cal C})$
having constraint density $c$ have resolution width at least $\eta n$. 

To accomplish this, we use the same strategy 
as in \cite{ben-sasson:resolution:width}: define for a unsatisfiable formula $\Phi$ a measure $\mu:Clauses \goesto {\bf N}$ (where $Clauses$ is the set of 
all possible disjunctions of literals from $\Var(\Phi)$, including 
the contradictory clause $\Box$) such that 
\begin{enumerate}
\item for every clause $C$ that appears in $\Cl(\Phi)$, $\mu(C)\leq 1$,
\item w.h.p.\ $\mu(\Box)$ is ``large''.
\item Infer that in any refutation there exists a clause $C$ with ``medium'' 
$\mu(C)$, and 
\item prove that if $\mu(C)$ is ``medium'' than the width of $C$ is
``large''. 
\end{enumerate}

As in \cite{ben-sasson:resolution:width}, define 

\[
\mu(C)=\min\{|\Xi|: \Xi\subseteq  \Phi, \Xi \models C\},
\]

where $\models$ is the logical entailment relation. In particular 
$\mu(\Box)$ is the size of the smallest unsatisfiable subformula of $\Phi$. 
 $\mu$ is subadditive, 
that is, for every clauses $C_{1}$ and $C_{2}$ that share a variable $x$ 
appearing with opposite signs in the two clauses, 
\[
\mu(res_{x}(C_{1},C_{2}))\leq \mu(C_{1})+\mu(C_{2}). 
\]

where $res_{x}(C_{1},C_{2})$ denotes the clause obtained by applying resolution to clauses 
$C_{1}$, $C_{2}$ with respect to variable $x$. 
It is clear that condition a) is satisfied.
As to b), the following is true: 

\begin{lemma}
There exists $\eta_{1}>0$ such that for any $c>0$, w.h.p.\
$\mu(\Box)\geq \eta_{1} n$, where  $\Phi$ is a random
instance of $\SAT({\cal C})$ having constraint density $c$.
\end{lemma}

%\iffalse
%%%%%%%%

\begin{proof}
In the proof of Theorem~\ref{sufficient:first-order} we have shown
that there exists $\eta_{0}>0$ such that w.h.p.\ any unsatisfiable
subformula of a given formula has at least $\eta_{0} n$
constraints. Therefore {\em any} formula $F$ made up of {\em clauses} from
the CNF-representation of constraints in $\Phi$, and which has
fewer than $\eta_{0} n$ clauses is satisfiable (since it is less tight 
than the conjunction of those constraints). 

The claim now follows by taking $\eta_{1} = \eta_{0}$.
\end{proof}
\qed
%\fi

The only (slightly) nontrivial step of the proof, which critically
uses the fact that constraints in ${\cal C}$ do not have
implicates of length one or two, is to prove that clause
implicates of subformulas of ``medium'' size have ``many''
variables. 

\begin{lemma}\label{expansion}
There exists $d>0$ and $\eta_{2}>0$ such that w.h.p.\ (when  $\Phi$ is a random
instance of $\SAT({\cal C})$ having constraint density $c$)
 every clause $C$ present in a refutation of $\Cl(\Phi)$ 
that satisfies $\frac{dn}{2}<\mu(C)\leq dn$ also satisfies  $|C|\geq
\eta_{2} n$.
\end{lemma}

\begin{proof}

Given a clause $C$, let $\Xi$ be a subformula of $\Phi$, having minimal size, such
that $\Xi \models C$. We claim:

\begin{lemma}\label{appear} 
For every constraint $P$ of $\Xi$ that contains $k-2$ private variables, 
at least one of these variables appears in $C$.
\end{lemma}

\begin{proof} 
Suppose there exists a constraint $D$ of $\Xi$ with at least $k-2$ private 
variables such that none of its private variables appears in $C$. Because of the minimality of $\Xi$ there exists an assignment $F$
that satisfies $\Xi \setminus \{D\}$ but does not satisfy $D$ or
$C$. Since $D$ has no implicates of size two, there exists an
assignment $G$, that differs from $F$ only on the private
variables of $D$, that satisfies $\Xi$. But since $C$ does not
contain any of the private variables of $D$, $F$ coincides with
$G$ on variables in $C$. The conclusion is that $G$ does not
satisfy $C$, contradicting the fact that $\Xi\models C$.
\end{proof} 
\qed

Now define $x(\cdot, \cdot)$ to be the function from \eqref{x-sparse} that 
describes the dependence of $x$ on $y$ and $c$. For a constant $\epsilon >0$ 
to be determined later, define 

\[
d = min(inf\{x(2/(2k-3+\epsilon),c)| c\geq c_{\SAT({\cal C})}\}, \eta_{1}).
\]

Since $\SAT({\cal C})$ has a sharp threshold, the first term of the minimum 
expression is, like $\eta_{1}$, strictly greater than zero. Therefore,
$d>0$.

\begin{lemma}\label{private}  
There exists constant $\eta_{2}>0$ such that w.h.p., when  $\Phi$ is a random
instance of $\SAT({\cal C})$ having constraint density $c$ and $W\subseteq \Phi$
 is a formula with at most $dn$ constraints,  $W$ contains at least $\eta_{2} n$ constraints each of which has at least $k-2$ private variables.  
\end{lemma} 
\begin{proof} 

To prove Lemma~\ref{private} we first need: 

\begin{lemma} \label{pr} 
Let $\epsilon>0$ be a constant. If $F$ is a formula with 
$c^{*}(F)\leq \frac{2}{2k-3+\epsilon}$ then for every
subformula $G$ of $F$, at least $(\epsilon/3)
|G|$ constraints of $G$ have  at
least $k-2$ private variables.
\end{lemma} 

\begin{proof} 
 Indeed,
since $c^{*}(G)\leq \frac{2}{2k-3+\epsilon}$, by an argument similar
to the one used in the proof of Lemma~\ref{1-transversal},   $v_{1}(G)\geq
(k-3+\epsilon)|G|$. Since constraints in $G$ have
arity $k$, at least $(\epsilon/3) |G|$ have
more than $k-3$ (i.e., at least $k-2$) private variables. 
\end{proof} 
\qed

Returning to the proof of Lemma~\ref{private}, 
choose $y=\frac{2}{2k-3+\epsilon}$ in
Lemma~\ref{sparsity:hypergraph} for $\epsilon>0$ a small enough
constant. Because of the definition of $d$, when $\Phi$ is a random
instance of $\SAT({\cal C})$ having constraint density $c$, w.h.p.\ formula   
$\Phi$ is $(d,y)$ sparse. Since $|W|\leq dn$, this easily implies the fact that 
\[
c^{*}(W) \leq \frac{2}{2k-3+\epsilon}. 
\]

Lemma~\ref{private} follows by applying Lemma~\ref{pr} to formula $W$
with $\eta_{2}=\epsilon/3$. Applying this result and Lemma~\ref{appear} to
formula $\Xi$ also concludes the proof of Lemma~\ref{expansion}. 

\end{proof} 
\end{proof} 
\qed

The proof of item 2. of
Theorem~\ref{implicates:first-order} now follows: since for any
clause $K$ in $\Cl(\Phi)$ we have $\mu(K)=1$, since
$\mu(\Box)>\eta_{1} n$ and since $0<d\leq \eta_{1}$, 
there indeed exists a
clause $C$ such that
\begin{equation}\label{intermediate}
\mu(C)\in [dn/2, dn].
\end{equation}

Indeed,
 let $C^{\prime}$ be a clause in the resolution refutation of $\Phi$ minimal with
the property that $\mu(C^{\prime})> dn$. Then at least one clause
$C$ of the two involved in deriving $C^{\prime}$ satisfies
\eqref{intermediate}.
Applying Lemma~\ref{expansion} we infer that the width of $C$ is at least 
$\eta_{2} n$. Using the size-width connection from\cite{ben-sasson:resolution:width} completes the proof of item 2. of Theorem~\ref{implicates:first-order}. 

\end{enumerate}

\end{proof}
\qedbox

\subsection{Threshold location and discontinuous spines}

Molloy \cite{molloy-stoc2002} has studied threshold properties of 
random constraint satisfaction problems, describing a technical 
property of the constraint set (called {\em very well-behavedness})
that is necessary for the existence of a sharp threshold. In 
\cite{istrate:sharp} we have shown that Molloy's {\em well-behavedness} condition is actually necessary and sufficient for boolean constraints (this has 
been independently proved by Creignou and Daud\'{e} \cite{creignou-daude-thresholds}). Thus we have 
completely characterized sets $C$ for which $\SAT({\cal C})$ has 
a sharp threshold. 

The well-behavedness condition has implication for the clause/variable 
ratio of minimally unsatisfiable formulas: it has to be larger than $1/(k-1)$. 
Furthermore, Molloy has shown that if the density of minimally unsatisfiable 
formulas is {\em bounded away from  $1/(k-1)$} (i.e., it satisfies 
the conditions of Theorem~\ref{sufficient:first-order}) then the location of 
the transition is {\em strictly larger than $\frac{1}{k(k-1)}$}. 

We have seen that the same density condition is sufficient to guarantee 
the discontinuity of the spine and exponential resolution complexity. 
A natural question therefore arises: is it
possible to relate the continuity (or discontinuity) of the spine to the
{\em location} of the phase transition ?

At first this does not seem to be possible: we have already encountered  
two problems that fail to satisfy the sufficient condition for a
discontinuous spine, random 2-SAT, for which the transition has been
proven to be of second order~\cite{scaling:window:2sat}, and random
1-in-$k$-SAT, for which a similar result
holds~\cite{istrate:1ink:sat}. Both have a threshold location 
strictly higher than Molloy's lower bound of $\frac{1}{k(k-1)}$. However, 
the most natural specification of the random model for the two problems 
involves applying constraints on both variables and their negations. 
For both problems the actual 
location of the threshold is {\em twice} the value given by Theorem 3 in
\cite{molloy-stoc2002}, at clause/variable ratio $\frac{2}{k(k-1)}$. This 
suggests that the following tempting intuitive picture might be accurate, 
at least in a more restricted setting:  

\begin{enumerate}

\item Problems with a continuous spine are ``2-SAT-like'', and have a
phase transition at constraint density $c^\cont_k= \frac{2}{k(k-1)}$. 
\item  Problems with a discontinuous spine
have a phase transition located at constraint density $c> c^\cont_k$. 
\end{enumerate} 

To obtain results that partly support the intuition above, we have to
modify the random model from Definition~\ref{model} to allow negated
variables. 

\begin{definition}
Let ${\cal C}$ be a set of constraint templates. The {\em closure of ${\cal C}$}, denoted $\overline{\cal C}$ is the set of 
constraints 
\begin{equation} \label{closure} 
\overline{\cal C}= \{C(x_{1}^{\epsilon_{1}}, \ldots, x_{k}^{\epsilon_{k}})\mbox{ }|\mbox{ }C\in {\cal C}\mbox{ and }\epsilon_{1}, \ldots, \epsilon_{k}\in \{\pm 1\}\}, 
\end{equation}   
where for a variable $x_i$ we define $x_i^{1}:=x_i$, $x_i^{-1}:=
\overline{x_i}$. 

Set ${\cal C}$ is {\em good} if $|\overline{\cal C}|=|{\cal C}| 2^{k}$, that is all elements on the right hand side of 
\eqref{closure} are distinct. 

\end{definition}

\begin{definition}\label{neg} 
Let ${\cal C}$ be a good set of constraint templates. 
Denote by $\SATneg({\cal C})$ the version of $\SAT({\cal C})$ that
generates a random formula by the following process: 

\begin{enumerate}
% \item $n$ is the number of variables.
% 
% \item $m$ is the number of constraints, chosen by the following
% procedure: first
\item Select, uniformly at random and with replacement,
$m$ hyperedges of the complete $k$-uniform hypergraph on $n$ variables.

\item For each hyperedge $e$, choose a random ordering $o_{e}$ of the variables
involved in it. 

\item Independently with probability 1/2 negate each variable appearing in $o_{e}$. 
\item Choose a random constraint template from  ${\cal C}$ and 
apply it to the ordered list of literals in $o_{e}$.
\end{enumerate}

% Again, when there is no need to be that specific we will drop the parameters 
% $n,m$ from our notation. 
% 
\end{definition} 

It is easy to see that problems such as $k$-SAT and 1-in-$k$-SAT can be
expressed using the framework of Definition~\ref{neg}.  The following
result shows that the intuition connecting the discontinuity of the
spine, resolution complexity and the location of the phase transition
does indeed have merit: a strenghtening of the condition guaranteeing
the existence of a discontinuous spine and exponential resolution
complexity also implies that the satisfiability threshold is
located at a value higher than $c^\cont_k$:

\begin{theorem}
Let ${\cal C}$ be a good set such that 
\begin{enumerate}
\item $\SATneg({\cal C})$ has a sharp threshold (the 
result in \cite{istrate:sharp} can be easily adapted to completely 
characterize such sets ${\cal C}$).   
\item  There exists $\epsilon > 0$ such that,  
for every minimally unsatisfiable formula $F$ whose constraints are drawn 
from template set ${\cal C}$, the ratio of the number of constraints in $F$ to the number of 
distinct literals (variables and negated variables) appearing in $F$ is 
at least $\frac{1+\epsilon}{k-1}$.                        
\end{enumerate} 
Then 

\begin{enumerate} 
\item There is a
constant $\delta >0$ such that random instances of $\SATneg({\cal C})$ 
with $m=cn$, where $c\leq \frac{2}{k(k-1)} (1+\delta)$, are 
satisfiable with probability $1-o(1)$. 
\item Problem $\SATneg({\cal C})$ has a discontinuous spine and exponential resolution complexity. 
\end{enumerate} 
\end{theorem}

\begin{proof} 
\begin{enumerate} 
\item 
Since  $\SATneg({\cal C})$ has a sharp threshold, it is sufficient to show 
that there exists a fixed constant $\eta >0$ such that the probability 
that a random formula is satisfiable is at least $\eta$. 

Suppose $m= cn$ with $c= \frac{2}{k(k-1)} (1+\delta)$, with $\delta > 0$ 
small enough. Define random model $\SATneg_2({\cal C})$ that is a
variant of $\SATneg({\cal C})$ as follows: 

\begin{enumerate} 
\item Choose a random $k$-uniform hypergraph $H$ with $m$ edges on the  vertex set (of cardinality $2n$) consisting of {\em variables and negated variables}.
\item For every edge $e\in H$ create a random permutation $o_{e}$ of its elements.  
\item Apply a random constraint template in ${\cal C}$ to variables in the 
ordered list $o_{e}$. 

\end{enumerate}

This model differs from the random model $\SATneg({\cal C})$ in that 
it allows for constraints that include a  $k$-tuple of literals
involving two opposite literals. 

Define $W$ to be the event that formula $\Phi$ contains some clause
involving two opposite literals. It is easy to see that the expected
number of such clauses in a random formula $\Phi$ is constant.
Therefore, with positive probability $\lambda >0$ in a random formula
generated according to  $\SATneg_{2}({\cal C})$, the bad event $W$
will {\em not} happen. 

Let $Z$ denote the event that a random formula with $m= cn$ clauses
generated according to the random model $\SATneg_2({\cal C})$ is satisfiable. 

Then the probability that a random formula in  $\SATneg({\cal C})$ is 
satisfiable is equal to $Pr[Z|\overline{W}]$. To show that this is bounded 
away from zero it is enough to prove that $Pr[Z]=1-o(1)$. 

The $k$-uniform hypergraph on the $2n$ nodes (variables 
and their negations) corresponding to choosing 
a random instance of $\SATneg_2({\cal C})$ is a random $k$-uniform hypergraph. Thus we want to show that 
a formula generated by first choosing such a random $k$-uniform hypergraph $H$, and then 
applying a random constraint template from ${\cal C}$ on the given
literals is w.h.p.\ satisfiable. 

The proof of this is entirely similar to a step in the proof of Molloy's Theorem 3 in \cite{molloy-stoc2002}, and 
amounts to showing that w.h.p.\ the hypergraph $H$ does not contain any hypergraph of high density, 
corresponding to the fact that minimally unsatisfiable subformulas have clause/variable density at least $\frac{1}{k-1} (1+\epsilon)$. 
Rather than repeating an argument that is presented in detail in that paper, we refer the reader to  \cite{molloy-stoc2002}. 

\item Since ${\cal C}$ is good, one can simply apply
Theorem~\ref{sufficient:first-order} to $\SAT(\overline{{\cal C}})$,
which is equivalent to problem $\SATneg({\cal C})$.
\end{enumerate} 
\end{proof} 
\qed

\section{Beyond random satisfiability: comparing the behavior of the backbone and spine}

In this section we investigate empirically the continuity of
the backbone for two graph problems, random three coloring (3-COL)
and the graph bipartition problem (GBP).  Both can be phrased as
decision or as optimization problems, in the same manner as $k$-SAT and
MAX-$k$-SAT.
% The latter problem is the decision counterpart 
% of the graph bisection problem in the same manner the $k$-SAT problem is 
% the decision counterpart of MAX-$k$-SAT. 

We consider a large number of instances of random graphs, of sizes up to
$n=1024$ and over a range of mean degree values near the threshold.
For each instance we determine the backbone fraction $f$.

Culberson and Gent~\cite{frozen:development} have shown experimentally
that the 3-COL spine fraction $f_{S_C}$, as defined in
Definition~\ref{spine-two}, exhibits a discontinuous transition. To be
consistent with this study, we use the backbone fraction $f_{B_C}$ from
the same definition. We employ a rapid heuristic called {\em extremal
optimization\/}~\cite{BoPe1}.  Although an incomplete procedure,
numerical studies~\cite{BoPe2} as well as testbed comparisons with an
exact algorithm ~\cite{Trick}, have shown that extremal optimization
yields an excellent approximation of $f_{B_C}$ around the critical
region (see~\cite{BoPe1} for further discussions, that, we believe,
convincingly support this assertion).  Fig.~1 shows $f_{B_C}$
as a function of mean degree. 

Culberson and Gent have speculated that at the 3-COL threshold, although
their spine is discontinuous, the backbone might be {\em continuous\/}.
The results in Fig.~1a suggest otherwise.  For 3-COL, $f_{B_C}$
does not appear to vanish above the threshold, indicating a
discontinuous large $n$ backbone~\cite{BoPe2}.

We next study the graph bipartion problem (GBP):

\begin{definition}
GBP is the following decision problem.  Given a (not necessarily
connected) graph $G$ with $n$ vertices, $n$ being an even number,
determine whether it can be partitioned into two edge-disjoint sets
having $n/2$ vertices each.
\end{definition} 
This problem cannot, strictly speaking, be cast in the setup of random
constraint satisfaction problems from Definition~\ref{model}, since not
every partition of vertices of $G$ is allowed. It can be cast to a
satisfiability problem (with variables associated to nodes, values
associated to each partition and constraint ``$x = y$'' associated to
the edge between the corresponding vertices) but we must add the
additional requirement that {\em all satisfying assignments contain an
equal number of ones and zeros}.  Thus the complexity-theoretic
observations of Section~\ref{main:section} do not automatically apply to
it.  We can, however, give a ``DPLL-like'' class of algorithms for GBP,
so the the hope of obtaining results similar to the previous ones is not
so far-fetched.

Let us investigate the continuity of the backbone/spine under the model
in Definition~\ref{spine-two}.  It is easy to see that the
constraint-based spine $S_C(G)$ of a GBP instance $G$ contains all edges
belonging to a connected component of size larger than $n/2$. Since the
GBP threshold takes place where the giant component becomes larger than
$n/2$, $f_{S_C}$ is discontinuous there.  On the other hand, the
backbone fraction $f_{B_C}$ (Fig.~1b) appears to remain
continuous, vanishing at large $n$ on both sides of the threshold. 

We have noted earlier that the discontinuity of $f_{B_C}$ is a stronger
property than the discontinuity of $f_B$.  Thus for 3-COL it follows
that the variable-based backbone is discontinuous as well.  By contrast,
it is not clear for GBP whether the variable-based backbone is
continuous: our preliminary experimental evidence is as yet
inconclusive. 

The results in Fig.~1b suggest that the spine and the backbone
can behave differently at the threshold, though they do not yet address
the question of whether the spine's discontinuity really has
computational implications for the decision problem's complexity.  After
all, unlike 3-COL, GBP can easily be solved in polynomial time by
dynamic programming. This situation is similar to that of XOR-SAT, where
a polynomial algorithm exists but the complexity of {\em resolution
proofs/DPLL algorithms} is exponential.  The class of ``DPLL-like''
algorithms that can solve GBP can no longer be simulated in a
straightforward manner by {\em resolution proofs}, however it can be
simulated using proof systems $Res(k)$ that are extensions of
resolution~\cite{krajicek:resk}. Some of the hardness results for
resolution extend to these more powerful proof systems, and in
\cite{istrate:descriptive} we investigate the extent to which our
present results apply to this class of proof systems.  These
preliminary results imply that, indeed, the discontinuity of the spine
{\em does} have computational implications for GBP.

\begin{figure}[H]
%\tabcapfont
\centerline{%
\begin{tabular}{c@{\hspace{2pc}}c}
\includegraphics[angle=-90, width = .45\linewidth]{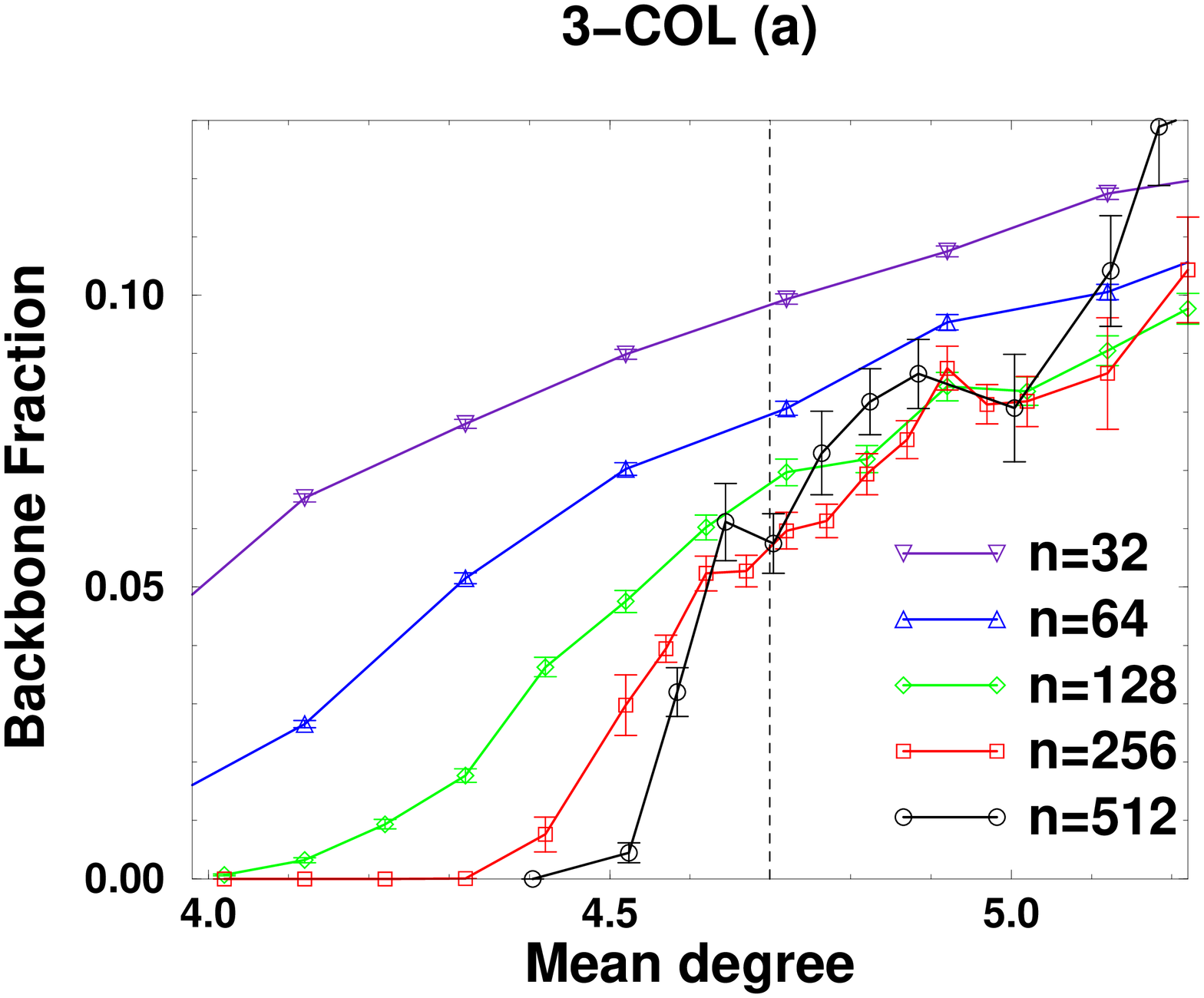} & 
\includegraphics[angle=-90, width = .45\linewidth]{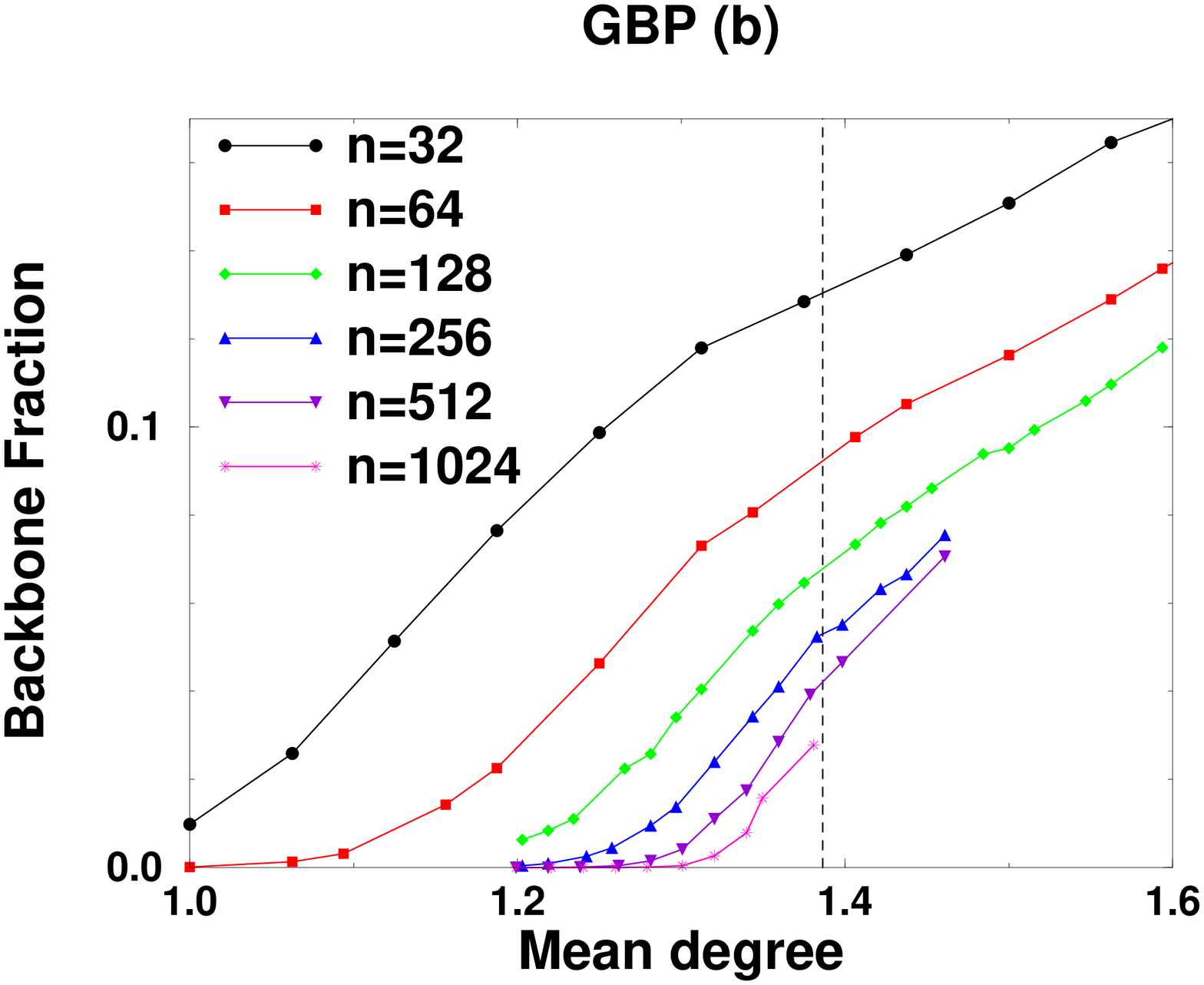}  \\
(a) 
3-COL~\cite{BoPe2} & (b) GBP. 
\end{tabular}}

\caption{
Plot of the estimated constraint-based backbone fraction $f_{B_C}$
on random graphs, as a function of mean degree $c$.  For
3-COL, the systematic error based on benchmark comparisons with random
graphs is negligible compared to the statistical error bars;
for GBP, $f_{B_C}$ is found by exact enumeration.
% For 3-COL we average over 2300, 500, and 280 instances for $n=32$, 64, and
% 128 at each value of $pn$.  For the GBP, we average over 3000, 350, and
% 100 instances for $n=50$, 100, and 200.
The thresholds
$c\approx4.70$ for 3-COL and $c=2\ln2$ for GBP are shown by dashed
lines.\label{mass}}
\end{figure}

\section{Discussion}

We have shown that the existence of a discontinuous spine
in a random satisfiability problem is often correlated with a
$2^{\Omega(n)}$ peak in the complexity of resolution/DPLL
algorithms at the transition point. The underlying reason is that the two phenomena
(the jump in the order parameter and the resolution complexity
lower bound) have common causes.

The example of random $k$-XOR-SAT
shows that a general connection between a first-order phase transition and
the complexity of the underlying decision problems is hopeless:
Ricci-Tersenghi et al.~\cite{zecchina:kxorsat}
have presented a non-rigorous argument using the replica method
that shows that this problem has a first-order phase transition, and
the following weaker result is a direct consequence of
Theorem~\ref{implicates:first-order}: 
 
\begin{proposition}
Random $k$-XOR-SAT, $k\geq 3$, has a discontinuous spine.
\end{proposition}

However, our results, as well as work in progress mentioned above,
suggest that the continuity/discontinuity of the spine is a predictor
for the complexity of the {\em restricted classes} of decision
algorithms that can be simulated by ``resolution-like'' proof systems.
Furthermore, experimental evidence in the previous section suggests that the
backbone and the spine do not always behave similarly.
Our analysis indicates that the spine, rather than the backbone, is the
order parameter to consider in studying the complexity of combinatorial
problems.

\section{Acknowledgments}

This work has been supported by the U.S.\ Department of Energy under
contract W-705-ENG-36, through the LANL LDRD program, and by grant
0312510 from the Division of Materials Research at the National Science
Foundation.

%\end{article}
\end{document}